\documentclass[twocolumn]{revtex4}
\usepackage{epstopdf}
\usepackage{graphicx}
\usepackage{dcolumn}
\usepackage{bm}

\newcommand{\gtsimeq}{\raisebox{-0.6ex}{$\,\stackrel
        {\raisebox{-.2ex}{$\textstyle >$}}{\sim}\,$}}

\def\gappeq{\mathrel{ \rlap{\raise.5ex\hbox{$>$}}
                      {\lower.5ex\hbox{$\sim$}}  } }
\def\lappeq{\mathrel{ \rlap{\raise.5ex\hbox{$<$}}
                      {\lower.5ex\hbox{$\sim$}}  } }

\begin{document}

\preprint{PRA}

\title{Matterwave Transport Without Transit}

\author{M. Rab$^1$, J.H. Cole$^{1,2}$, N.G. Parker$^1$, A.D. Greentree$^{1,2}$, L.C.L. Hollenberg$^{1,2}$ and A.M. Martin$^1$}

\address{$^1$School of Physics, University of Melbourne, Parkville,
Victoria 3010, Australia.}
\address{$^2$Centre for Quantum Computer Technology, School of Physics, University of Melbourne, Parkville,
Victoria 3010, Australia.}
\date{\today}

\begin{abstract}

Classically it is impossible to have transport without transit,
i.e., if the points one, two and three lie sequentially along a path
then an object moving from one to three must, at some point in time,
be located at two. However, for a quantum particle in a three-well
system it is possible to transport the particle between wells one
and three such that the probability of finding it at any time in the
classically accessible state in well two is negligible. We
consider theoretically the analogous scenario for a Bose-Einstein
condensate confined within a three well system.  In particular, we
predict the adiabatic transportation of an interacting
Bose-Einstein condensate of 2000 $^7$Li atoms from well one to well
three  without transiting the allowed intermediate region. To an
observer of this macroscopic quantum effect it would appear that, over a timescale of the order of
$1$s, the condensate had transported, but not transited, a
macroscopic distance of $\sim 20\mu$m between wells one and three.

\end{abstract}

\maketitle
The system under consideration is schematically
shown in Fig. 1(a), where a three-dimensional harmonic trap is split
into three regions via the addition of two parallel repulsive
Gaussian potentials. With the Bose-Einstein condensate (BEC) [blue object in Fig. 1(a)], initially in well 1, we show how it is
possible, through adiabatic changes to the tunneling rates between the wells, to transport it into well
 3 with minimal (ideally zero) occupation of the intervening well. This effect as a function of time is shown in Fig. 1(b), where an interacting BEC of 2000 $^7$Li atoms
is transported from well 1 to well 3 over a timescale of $\sim
1$s, with less than $1\%$ atoms occupying well 2 at any particular
time. As such it appears that the BEC is transported from well 1
to well 3 without transiting through well 2.

This effect of transport without transit (TWT) can be likened to the lay concept of teleportation. However, although TWT relies on quantum control of the global BEC state and associated tunneling matrix elements, it is quite distinct from the quantum definition of teleportation \cite{Bennett}.
In the TWT of a BEC we describe the many body system in a time dependent mean-field approximation. As such the {\it wavefunction} used to describe the condensed state is a classical field and can not describe such properties as entanglement and hence quantum teleportation.

\begin{figure}
\centering
\includegraphics[width=8.0cm]{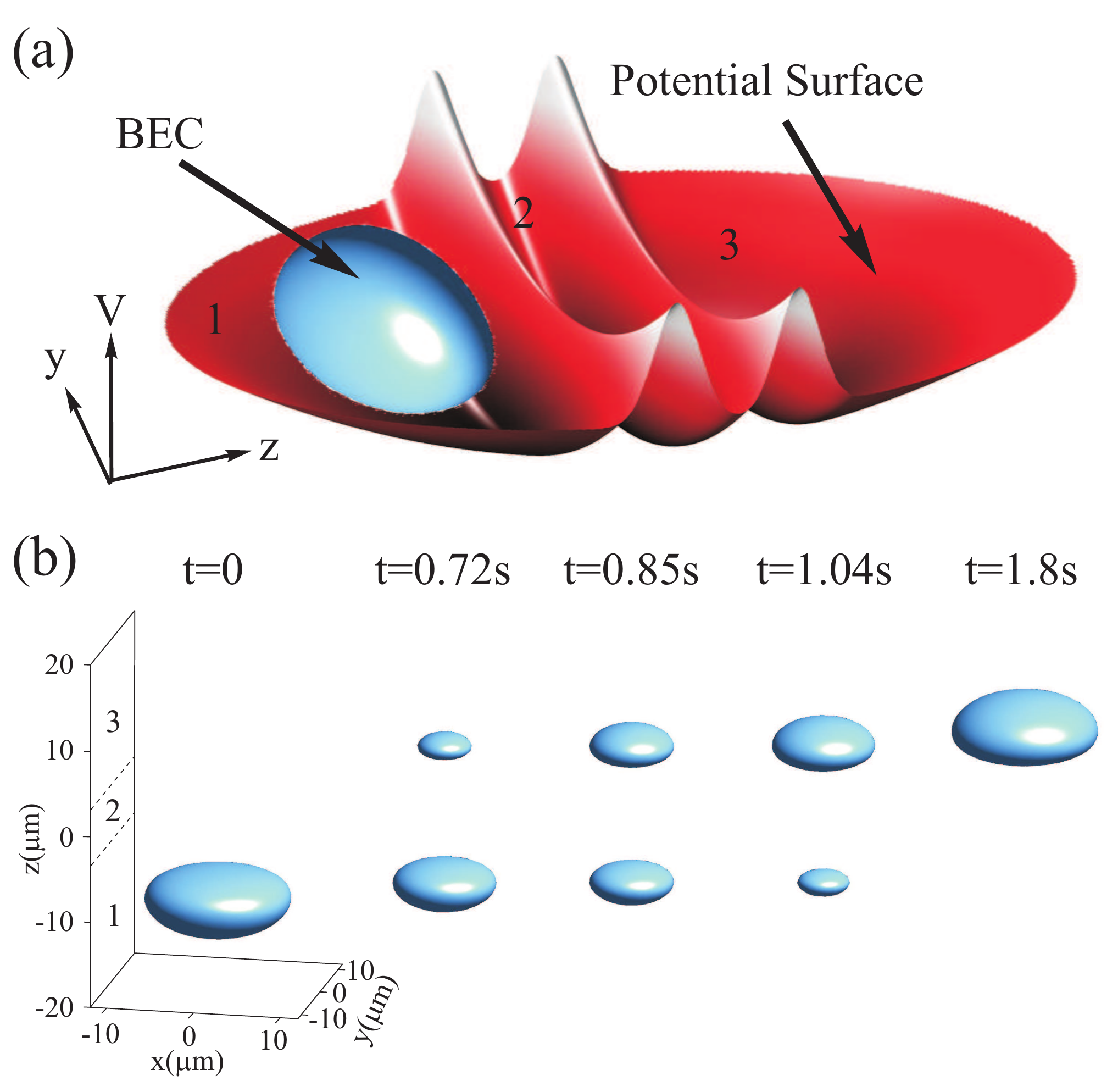}
\caption{Transport without transit (TWT) of a BEC in a
three well system. (a) Schematic representation of our system at t=0
in the $(z-y)$ plane. Two parallel, repulsive Gaussian barriers
embedded in an ambient harmonic trap divide the system into three
wells, with the BEC initially occupying well one. (b) Isosurface
plots of atomic density ($n_{\rm iso}=0.1 n_0$, where $n_0$ is the
initial peak density of the BEC) showing the adiabatic
transportation of a BEC of 2000 $^7$Li atoms over a distance of $20
\mu$m (simulated using the 3D Gross-Pitaevskii equation).}
 \vspace{-0.5cm}
\end{figure}

The ideas underpinning the protocol for TWT stem from Stimulated
Raman Adiabatic Passage (STIRAP)
\cite{Oreg,Kuklinski,Gaubatz,Bergmann}. STIRAP is a robust optical
technique for transferring population between two atomic
states, $|1\rangle$ and $|3\rangle$, via an intermediate
excited state, $|2\rangle$. Using off-resonant pulses to couple
states $|1\rangle$ to $|2\rangle$ and $|2\rangle$ to
$|3\rangle$, characterised by coupling parameters $K_{12}$ and
$K_{23}$, and such that $K_{23}$ precedes and overlaps $K_{12}$, the
population can be adiabatically transferred from state $|1\rangle$
to $|3\rangle$. Population transfer is achieved via a
superposition of states $|1\rangle$ and $|3\rangle$ with
the occupation of state $|2\rangle$ strongly suppressed. These
techniques are  used in quantum optics for coherent internal state
transfer \cite{Bergmann,Weitz,Wynar,Winkler} and have been proposed for applications in three channel optical waveguides \cite{Longhi} while an analogous approach has been proposed for state transfer from one atom laser beam to another \cite{Murry}.   Recently this
protocol has been proposed to transport single atoms \cite{Eckert,Deasy}, Cooper pairs \cite{Brandes} and electrons
\cite{Zhang,CTAP1,Fabian,CTAP2}.  Here we
extend these ideas to the transport of dilute gas BECs containing
thousands of atoms.

In this article we elucidate the properties of the three-well
system by first considering a three-mode approximation
\cite{Graefe,Wang,Liu}, where the form of the potential is not
important.  We then employ the  mean-field
Gross-Pitaevskii equation (GPE) to qualitatively describe the BEC
dynamics and consider experimental scenarios in which to realise macroscopic
matter-wave TWT.

Reducing our three-well system, shown in Fig. 1(a), such that each well is described by a single mode basis
\cite{Graefe,Wang,Liu}, $\Psi_i$, enables its properties to be described via
\begin{eqnarray}
\Psi(t)=\psi_1(t)\Psi_1+\psi_2(t)\Psi_2+\psi_3(t)\Psi_3
\end{eqnarray}
where
\begin{eqnarray}
\frac{i}{\Omega_{{\rm max}}}\frac{\partial}{\partial
t}\left(
    \begin{array}{c}
      \psi_1 \\
      \psi_2 \\
      \psi_3 \\
    \end{array}
  \right)=\left(
            \begin{array}{ccc}
              U_1 & -K_{12} & 0 \\
              -K_{12} & U_2 & -K_{23} \\
              0 & -K_{23} & U_3 \\
            \end{array}
          \right)\left(
    \begin{array}{c}
      \psi_1 \\
      \psi_2 \\
      \psi_3 \\
    \end{array}
  \right). \label{Mode}
\end{eqnarray}

The amplitude of each mode is expressed as $\psi_i=\sqrt{N_i}e^{i\theta_i}$,
where $N_i$ and $\theta_i$ are the occupation and phase of the $i$th
mode, respectively ($i=1,2,3$). The system is normalised such that
$\sum_{i=1}^3 N_i(t)=N_T$, where $N_T$ is the total number of atoms
in the system.  The parameters $K_{12}$ and $K_{23}$ describe the
wavefunction overlap, and hence tunneling rate, between wells $1$
and $2$, and $2$ and $3$, respectively. Furthermore, the
dimensionless on-site interaction energy per particle is
given by $U_i=E_i^0+g_m N_i/N_T$, where $E_i^0$ is the groundstate
energy of well $i$ and $g_m$ is a dimensionless parameter describing
the nonlinear atomic interactions within the system \cite{note}.

The modulation of the wavefunction overlaps $K_{12}$ and $K_{23}$
controls a transfer of atoms between the wells.  We assume these
parameters  vary with time as,
\begin{eqnarray}
K_{12}(t)&=& \sin^2[\pi t/(2t_{{\rm p}})], \nonumber \\
K_{23}(t)&=& \cos^2[\pi t/(2t_{{\rm p}})] \label{rates}
 \end{eqnarray}
where $t_{{\rm p}}$ is the total pulse time and the maximum
tunneling rate is defined by $\Omega_{{\rm max}}$.  We employ this protocol due to it robustness against non-linear effects arising from the inter-atomic interactions at $t=0$ and $t=t_{{\rm p}}$.

In the limit $t_{{\rm p}}\rightarrow \infty$ and for $g_m=0$ the
evolution of the modes are given by \cite{CTAP1},
\begin{eqnarray}
D_+&=&\sin\Theta_1 \sin \Theta_2 \, \Psi_1+\cos \Theta_2 \, \Psi_2 + \cos\Theta_1 \sin \Theta_2 \, \Psi_3  \nonumber \\
D_-&=&\sin\Theta_1 \cos \Theta_2 \, \Psi_1- \sin \Theta_2 \, \Psi_2+ \cos\Theta_1 \cos \Theta_2 \, \Psi_3 \nonumber \\
D_0&=&\cos\Theta_1 \, \Psi_1 - \sin \Theta_1 \, \Psi_3
\end{eqnarray}
where
\begin{eqnarray}
\Theta_1&=&\arctan\left[K_{12}/K_{23}\right] \\
\Theta_2&=&\frac{1}{2} \arctan \left[ \left(
2\sqrt{K_{12}^2+K_{23}^2}\right)/E_2^0\right].
\end{eqnarray}
The corresponding mode-energies are shown in Fig.~2(a). For an
initial state where all of the atoms are in well 1 the system
adiabatically follows the green/middle line. This corresponds to the
passage of atoms from well 1 to well 3, with a heavily
suppressed occupation of well 2, as shown in Fig. 2(b), corresponding to TWT.
\begin{figure}
\centering
\includegraphics[width=8.0cm]{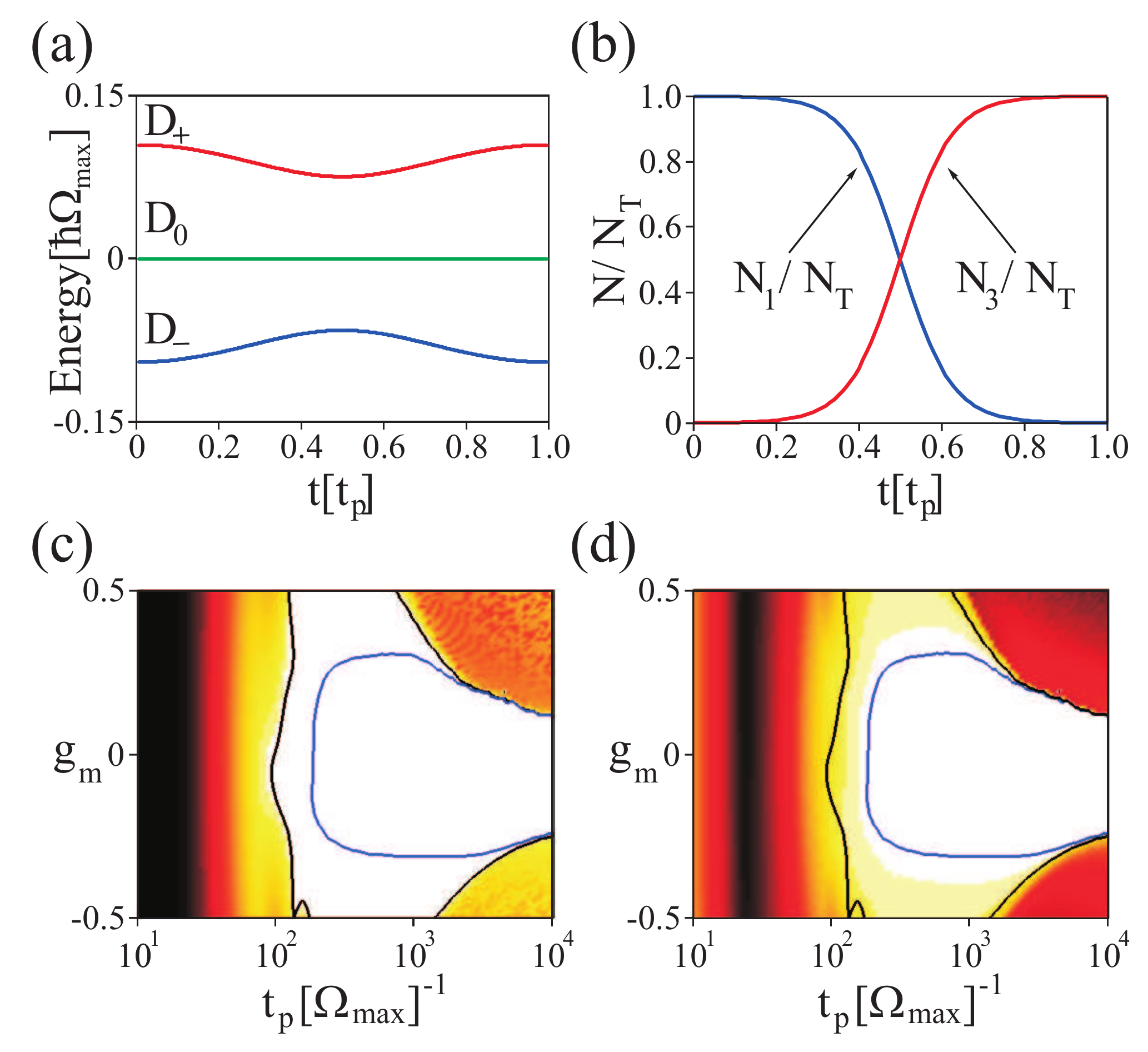}
\caption{Dynamics of the system according to the three-mode
analysis. (a) Energies of the eigenstates $D_+$, $D_0$ and $D_-$ of
the non-interacting ($g_m=0$) system. (b) Evolution of $N_1(t)/N_T$
and $N_3(t)/N_T$ for $g_m=0$ and $t_{{\rm p}} \rightarrow \infty$.
(c) $N_3(t=t_{{\rm p}})/N_T$ as a function of $g_m$ and $t_p$, with white and black representing $N_3(t=t_{{\rm
p}})/N_T=1$ and $0$ respectively. (d) max$[N_2(t)/N_T]$ as a function of $g_m$
and $t_{{\rm p}}$, with white and black representing $N_3(t=t_{{\rm
p}})/N_T=0$ and $1$ respectively.   In (c) and (d) the solid black and blue/grey
curves represent $N_3(t=t_{{\rm p}})/N_T=0.99$ and
max$[N_2(t)/N_T]=0.01$ respectively, and the region bounded by both
corresponds to high fidelities $\epsilon\geq 0.99$.  We have assumed
$E_1^0=E_3^0=0$ and $E_2^0=0.1\hbar \Omega_{{\rm max}}$ throughout.}
 \vspace{-0.5cm}
\end{figure}

Figures 2(a,b) are in an ideal limit where the atomic interactions
are zero ($g_m=0$) and the time over which the pulses were applied was
large ($t_p \rightarrow \infty$). However, for a
realistic system it is instructive to examine how this ideal picture
changes as the pulse times ($t_{{\rm p}}$) are reduced and
interactions are included \cite{Graefe}.  It is possible to parameterise the
evolution of the system via the fraction of atoms in well 3 at the end of the protocol, $N_3(t_{{\rm p}})/N_T$, and the maximum number of atoms occupying well 2 during the protocol, ${\rm max}[N_2(t)/N_T]$. These quantities
are mapped out in Figs. 2(c) and (d), respectively, as a function of
the strength of the nonlinear interactions $g_m$ and the pulse time
$t_p$, with efficient TWT occurring in the white regions. Defining a fidelity, $\epsilon$, for TWT through
$N_3(t_{{\rm p}})/N_T>\epsilon$ and ${\rm max}[N_2(t)/N_T] <
1-\epsilon$ we find that to achieve $\epsilon \geq 0.99$ we require
that $|g_m| < E_2^0-E_1^0$ and $t_{{\rm p}} \Omega_{{\rm max}}
\gtsimeq
 400$. In the absence of nonlinear interactions ($g=0$) the condition for
$t_{{\rm p}} \Omega_{{\rm max}} > 400$ comes from the adiabatic
limit of the system and is governed by the energy difference between
the groundstate energies of the wells. As noted by Graefe {\it et
al.} \cite{Graefe}, the introduction of nonlinear interactions
introduces new nonlinear ``eigenstates'', which can inhibit adiabatic transfer. We note that
in their approach they considered a Gaussian tunneling scheme.
However, the protocol which we employ [Eq. (\ref{rates})] is much
more robust to non-linear  effects, since the energies of the additional
non-linear states are not close to the dark state mediating the
transfer.

The mode analysis presented above gives a qualitative description of
adiabatic transport for a three well system. To investigate TWT
quantitatively for realistic scenarios the GPE is employed.
The GPE mean-field model has had great success in describing the
dynamics of BECs, e.g. the formation of vortex lattices in rotating
BECs \cite{Rapid,Nick}, the quantum reflection of BECs off silicon
surfaces \cite{Pasquini,PRL_reflection}, the breakdown of Bloch
oscillations of BECs in optical lattices \cite{PRL_Bloch,PRA_Bloch}
and the formation of bright solitary waves in attractive BECs
\cite{JPhysB}.

The GPE model is valid in the limit of zero temperature and
describes the BEC by a macroscopic order parameter, or
``wavefunction'', $\Psi({\bf r},t)$ which represents the mean-field
of the Bose-condensed atoms. This macroscopic wavefunction can be
expressed as $\Psi({\bf r},t)=\sqrt{n({\bf r},t)}\exp[i\theta({\bf
r}, t)]$, where $n({\bf r},t)$ is the atomic density and
$\theta({\bf r}, t)$ is a macroscopic phase. The
evolution of the wavefunction $\Psi({\bf r},t)$ is described by the
GPE,
\begin{eqnarray}
i \hbar \frac{\partial \Psi({\bf r},t)}{\partial t}= \left[
-\frac{\hbar^2}{2m}\nabla^2 + V({\bf r},t)+g \left|\Psi({\bf r},t)
\right|^2 \right]\Psi({\bf r},t).\nonumber \\
&&
 \label{GPE}
\end{eqnarray}
Here the nonlinear coefficient is given by $g=4\pi\hbar^2 a/m$,
where $a$ is the s-wave scattering length that characterises the
atomic interactions in the BEC.  We assume a trapping potential
$V({\bf r},t)$ of the form,
\begin{eqnarray}
&&V({\bf r},t)=\frac{m}{2}\left[\omega_\perp^2(x^2+ y^2)+\omega_z^2
z^2\right] \nonumber\\
&+&
V_{12}(t)\exp\left[-\frac{\left(z+z_0\right)^2}{2\sigma^2}\right]
+V_{23}(t)\exp\left[-\frac{\left(z-z_0\right)^2}{2\sigma^2}\right].\nonumber \\
&& \label{potential}
\end{eqnarray}
The first term defines the cylindrically-symmetric parabolic trap
with radial and axial trap frequencies $\omega_\perp$ and
$\omega_z$, respectively. The second and third terms represent the
repulsive Gaussian barriers, positioned at $z=\pm z_0$, with width
$\sigma$ and time-dependent amplitudes $V_{12}(t)$ and $V_{23}(t)$.
Such barrier potentials can be induced by the optical dipole force
from two parallel blue-detuned laser beams \cite{Ketterle} or using
magnetic fields on an atom chip \cite{chip}. The
barrier amplitudes, which can be varied by modifying the laser intensity or the atom chip currents,  controls the tunnelling rates between the
neighboring wells.

Recall that in the mode analysis we employed tunnelling rates
which initially had opposing values [Eq.~(\ref{rates})].  Due to
difficulties in initialising such a state in an experiment and
simulation, we begin our simulations with barriers of identical
height $V_{{\rm max}}$, and therefore initially identical tunnelling
rates. The exact functional form of $V_{12}(t)$ and $V_{23}(t)$ that
we employ can be found in the Methods section.  Over time, the barriers are smoothly
lowered to a minimum value $V_{\rm min}$ before being increased back
to $V_{\rm max}$, such that the tunnelling rate variation  approximates a  Gaussian. Importantly, there is a time delay of size $\tau$
between the pulsing of $V_{12}(t)$ and $V_{23}(t)$. When $V_{12}(t)$
is pulsed before $V_{23}(t)$, we term this the {\em intuitive}
protocol, and when $V_{12}(t)$ is pulsed after $V_{23}(t)$, we term
this the {\em counter-intuitive} protocol. Only the latter protocol
is capable of producing TWT
 and so we
concentrate on this.

\begin{figure}
\centering
\includegraphics[width=8.0cm]{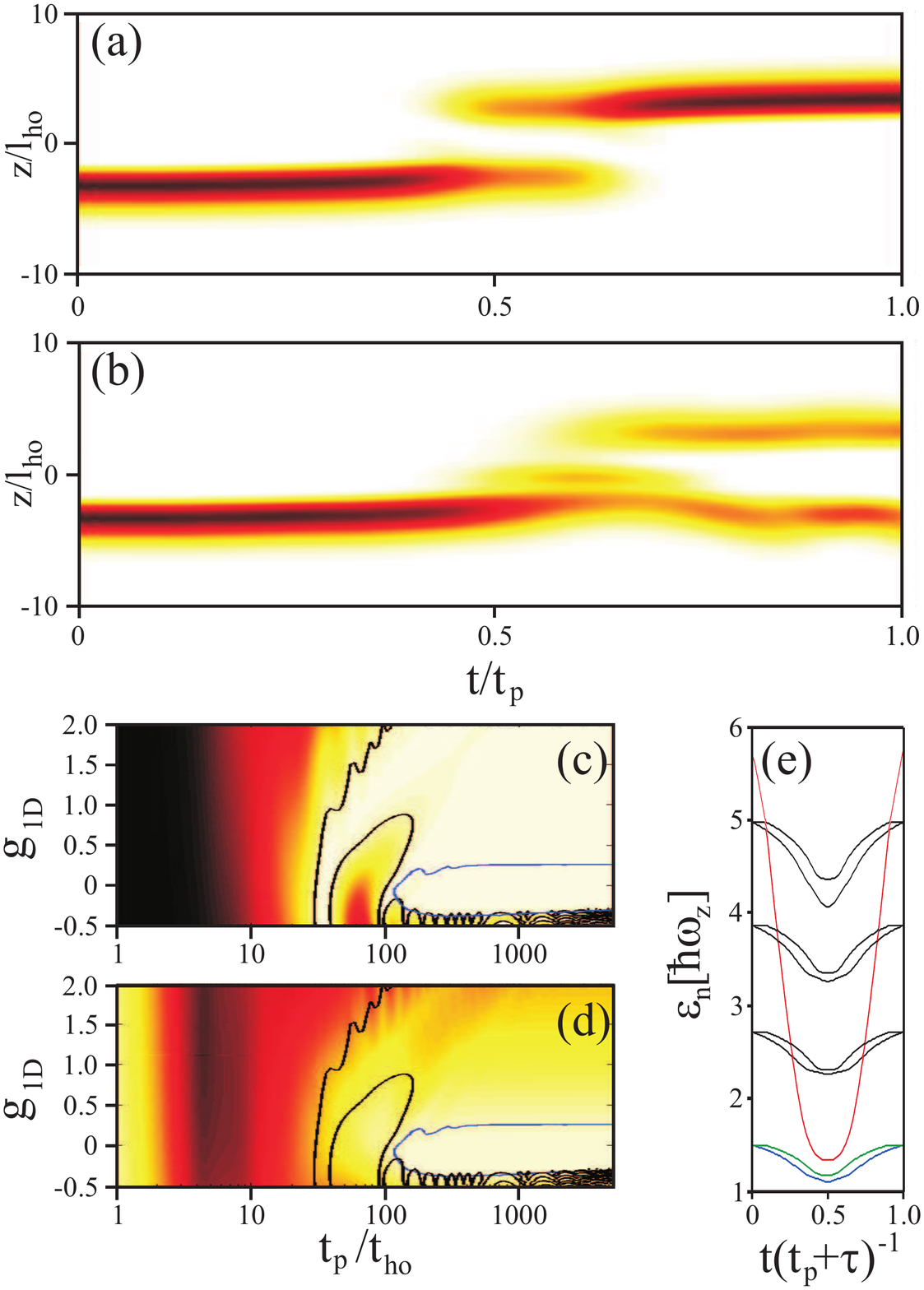}
\caption{Dynamics of the system according to the 1D GPE.  (a) Carpet
plot showing the evolution of condensate density (dark=high density,
light=low density) for an effective 1D interaction parameter of
$g_{1D}=0.31$ and a pulse time of $t_{\rm p}=1000t_{\rm ho}$. (b)
Same as (a) but for a reduced pulse time of $t_{{\rm p}}=14t_{{\rm
ho}}$. Note the breakdown of the adiabatic transfer.  (c)
$N_3(t=t_{{\rm p}})/N_T$ as a function of $g_{{\rm 1D}}$ and
$t_{{\rm p}}$, with white and black representing
$N_3(t=t_{{\rm p}})/N_T=1$ and $0$ respectively.  (d) max$[N_2(t)/N_T]$ as a
function of $g_{{\rm 1D}}$ and $t_{{\rm p}}$, with white and black
representing max$[N_2(t)/N_T]=0$ and $1$ respectively.  In (c) and (d) the
solid black and blue curves represent $N_3(t=t_{{\rm p}})/N_T=0.95$
and max$[N_2(t)/N_T]=0.05$ respectively, and the region bounded by
both corresponds to high fidelities $\epsilon\geq 0.95$. (e) The evolution of the first nine
energy eigenvalues ($\epsilon_n$) as a function of time ($g_{{\rm
1D}}=0$), with the blue, green and red curves corresponding to the
$D_-$, $D_0$ and $D_+$ mode energies respectively. In all cases $V_{{\rm min}}=5\hbar \omega_z$, $V_{{\rm
max}}=10^4\hbar \omega_z$, $\sigma=0.16l_{{\rm ho}}$ and $z_0=0.48l_{{\rm ho}}$.}
 \vspace{-0.5cm}
\end{figure}

The difference between the pulsing schemes used for the mode
analysis and GPE simulations is seen by comparing the adiabatic
evolution of the eigenstates in the two approaches, Figs. 2(a) and
3(e). Despite the qualitatively different TWT protocols for the
mode analysis and GPE simulations, we see qualitative agreement
in the regions of high fidelity. This suggests that the method of
transporting BECs is not particularly dependent on the exact form of
$V_{12}(t)$ [$K_{12} (t)$] and $V_{23}(t)$ [$K_{23}(t)$], as expected for an adiabatic protocol. This has been verified through the study of several different functional forms for $V_{12}(t)$ [$K_{12}(t)$] and $V_{23}(t)$ [$K_{23}(t)$] which all produce qualitatively similar results.

The one-dimensional equivalent of the GPE can be solved numerically
with relative ease and so we consider this limit first.  Employing
harmonic oscillator units, as outlined in the Methods section, we consider a system defined by $\sigma=0.16 l_{{\rm
ho}}$, $z_0=3\sigma$ and $\tau=t_{{\rm p}}/10$.

Figures 3(a,b) show the evolution of the condensate density
$|\Psi(z)|^2$ for different time pulses but the same remaining
parameters ($g_{{\rm 1D}}=0.31$, $V_{{\rm min}}=5\hbar \omega_z$ and
$V_{{\rm max}}=10^4\hbar \omega_z$). In Fig. 3(a) a large time pulse
of $t_{{\rm p}}=1000t_{{\rm ho}}$ leads to efficient TWT, with the
BEC moving smoothly from well 1 to well 3 with a minimal
occupation of well 2. In Fig. 3(b), however, a significantly
reduced pulse time of $t_{\rm p}=14t_{{\rm ho}}$ breaks the
adiabaticity of the process and causes inefficient transfer, with a
significant population in well two.

In Figs. 3(c,d)  $N_3(t_{{\rm p}})/N_T$ and ${\rm max}[N_2(t)/N_T]$
are plotted as a function of $t_{{\rm p}}$ and $g_{{\rm 1D}}$, in
analogy to the three-mode results in Figs. 3(c,d). The curves in
Figs. 3(c) and (d) denote $N_3(t^{{\rm p}})/N_T=\epsilon$ (black)
and ${\rm max}[N_2(t)/N_T]=1-\epsilon$ (blue) for $\epsilon=0.95$,
according to the GPE. These results are qualitatively similar to the mode analysis and show a large region of the parameter space where efficient TWT can occur.

\begin{figure}
\centering
\includegraphics[width=8.0cm]{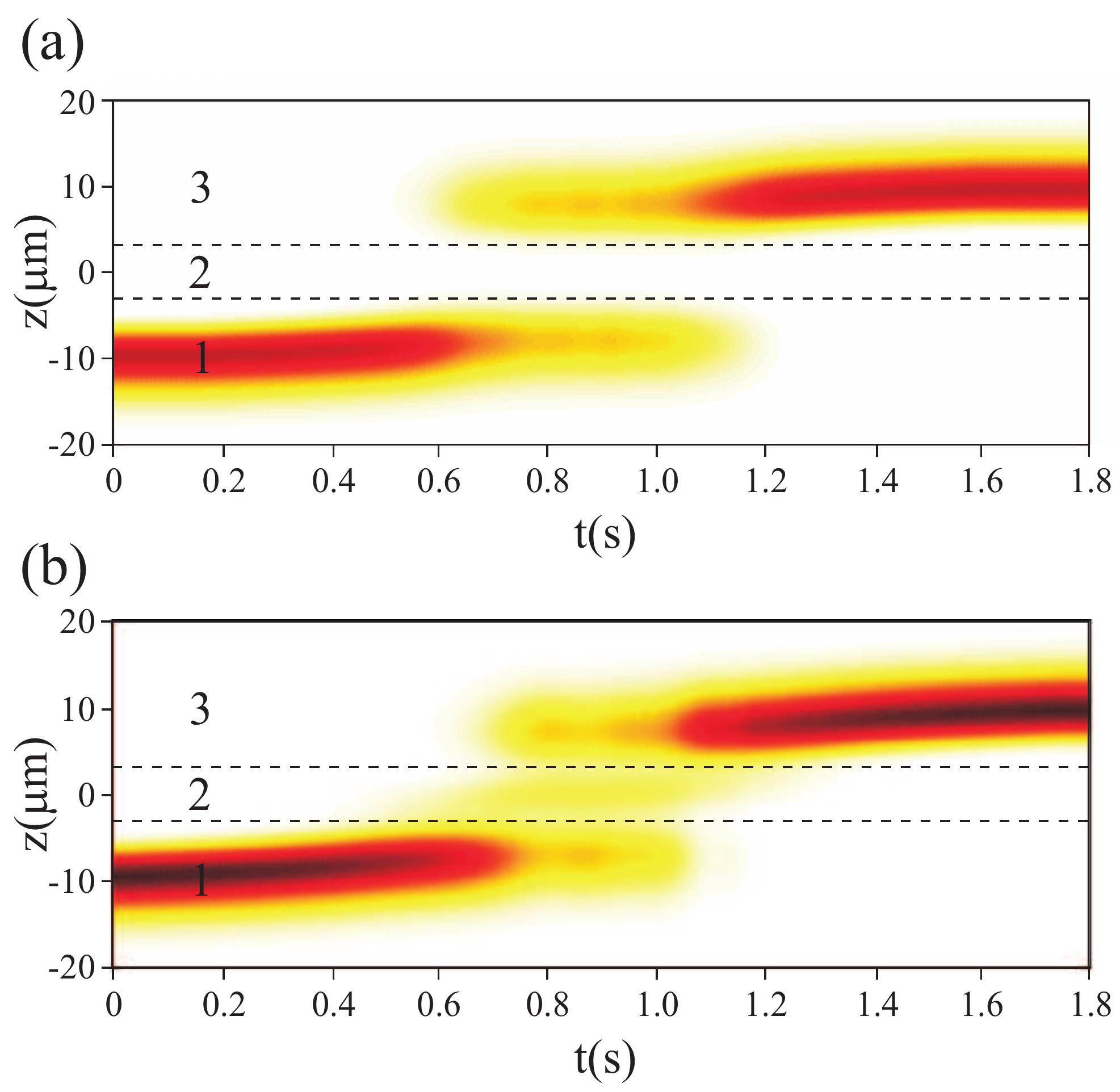}
\caption{Demonstration of TWT for a realistic BEC of $2000$ $^7$Li
atoms and attractive interactions $a=-0.2$nm. (a) Carpet plot
showing the evolution of the radially-integrated axial density for
the {\em counter-intuitive} protocol (defined by Eqs. (\ref{V23})
and (\ref{V12})).  (b) Same as (a) but for the {\em intuitive}
protocol [$V_{23}(t) \rightarrow V_{12}(t)$ and $V_{23}(t)
\rightarrow V_{12}(t)$] giving an appreciable population in the middle middle well. In each of the plots the horizontal dashed
lines correspond to the center of the Gaussian barriers $\pm z_0$.
We assume $\omega=2\pi\times 40$Hz, $\sigma=1\mu$m, $z_0=3\mu$m,
$V_{{\rm max}}=100\hbar \omega$ and $V_{{\rm min}}=5\hbar \omega$.}
 \vspace{-0.5cm}
\end{figure}

We now consider the possibility of producing efficient TWT in
a realistic BEC system.  We performed simulations of the
full 3D GPE.  Since strong nonlinear interactions suppress TWT we
focus on a system with weak interaction strength, i.e. a small
s-wave scattering length $a$ and low atom number $N_T$. Our
 simulations are based on recent $^7$Li soliton experiments
\cite{Khaykovich,Strecker}.  These experiments have two key
advantageous features.  Firstly, the experiments worked with low
atom number, with typically several thousand atoms in the
condensate. Secondly, the experiments employed a Feshbach resonance
to control the s-wave scattering length and, indeed, this allowed
the use of a low attractive scattering length of the order of
$a=-0.1$nm. This means that in principle similar experiments could
probe high fidelity parameter space of $g$ and $t_p$.

We consider $N_T=2000$ and $a=-0.2$nm, and
realistic parameters for our trapping system:
$\omega_r=\omega_z=2\pi \times 40$Hz, $\sigma=1\mu$m, $V_{{\rm
max}}=100\hbar\omega$, $V_{\rm min}=5\hbar\omega$ and $z_0 = 3\mu$m.
Initially we consider the condensate dynamics under the
counter-intuitive protocol and for a pulse time of $t_{\rm
p}=400\omega^{-1}=1.6$s and a pulse delay of $\tau=0.16$s. These
dynamics are presented in Fig. 1(b) as chronological frames of an
isosurface of the BEC density and in Fig. 5(a) as a carpet plot of
the radially-integrated axial density. For these realistic parameters we clearly see efficient
TWT, i.e. the $2000$ $^7$Li atoms are adiabatically transported a
distance of approximately $20 \mu$m with negligible
occupation of well 2. Crucially, the timescale for this process is
just under $2$s, which is the lifetime of such condensates
\cite{Strecker}. These results have a fidelity $\epsilon=0.985$,
which is limited by the maximum occupation of well 2 during the
transfer. Up until now we have defined well 2 as the spatial
region $[-z_0,z_0]$ for simplicity. However, it may be more
appropriate to define well 2 as the classically-allowed region,
i.e. the region between $-z_0$ and $z_0$ where the chemical
potential of the initial state is less than $V({\bf r},t)$. Under
this definition we find that the maximum atom number at any given
time that occupies the classically-allowed region is less than $1\%$
of $N_T$, giving a fidelity of $\epsilon>0.99$.

We have also simulated the dynamics of this system for the {\em
intuitive} protocol.  Recall that this corresponds to when the first
barrier is pulsed before the second barrier. These dynamics are
presented in Fig. 5(b), which shows the evolution of the
radially-integrated axial density.  Under this protocol we clearly
see the macroscopic occupation of well 2 during the transfer.
Indeed, at a single time during these dynamics over $15\%$ of the
atoms reside in the classically allowed region in well 2, two
orders of magnitude larger than for the counter-intuitive protocol.
This demonstrates that a straightforward experimental
confirmation of TWT is to reverse the pulses and compare the
condensate density in the middle well half way through the pulse sequence.

In conclusion we propose a novel protocol for the transport of BECs in three-well systems. This protocol enables the   adiabatic
transport a macroscopic BEC such that the transient occupation
of the intermediate well is heavily suppressed: transport without transit. In particular, we
have shown that this works within both a three-mode approach and a
meanfield approximation, where all of the modes of the mean-field
system are considered. We have mapped out the parameter space for
which we expect transport without transit to occur. Specifically, we
have demonstrated the transport-without-transit of an interacting
BEC of $2000$ $^7$Li atoms a macroscopic distance of $20\mu$m over a
timescale of $1.8$s. This phenomenon is not only of interest from
the view point of testing the {\it wave} nature of a dilute gas
Bose-Einstein condensate, but also paves the way for a new method of
control in atom optical devices. Future extensions to this work include the examination of non-meanfield effects, such as quantum fluctuations \cite{fluctuations}, and the consideration for systems with more than three wells \cite{CTAP3}.

The authors acknowledge useful discussions with Simon Devitt and David Jamieson. This work is funded by the
Australian Research Council. Additionally JHC, ADG and LCLH are
supported by the US National Security Agency (NSA), Advanced Research and Development Activity (ARDA) and the Army Research Office (ARO) under Contract Nos. W911NF-04-1-0290.

\section{Methods}
To induce an approximately Gaussian modulation of the tunnelling rates
\cite{CTAP1,Graefe} we need a functional form for the barrier
heights of,
\begin{eqnarray}
V_{23}(t)= \left\{
            \begin{array}{c}
              16\left(V_{{\rm max}}-V_{{\rm
min}}\right)\left(\frac{t}{t_{{\rm p}}}-\frac{1}{2}\right)^4+V_{{\rm
min}} \, \, \, \, \, \, \, \, \, \,  t<t_{{\rm p}} \\
               V_{{\rm max}} \, \, \, \, \, \, \, \, \, \, \, \, \, \, \, \, \, \, \, \, \, \, \, \, \, \, \, \, \, \, \, \, \, \, \, \, \, \, \, \, \, \, \, \, \, \, \, \, \, \, \, \, \, \, \, \, \, \, \, \, \, \, \, \, \, \, \, \, \, \, \, \, \, \, \, \, \, \, \, \, \, \, \, \, \, \, \,  t \ge
               t_{{\rm p}}\\
            \end{array}
          \right.
          \label{V23}
\end{eqnarray}
\begin{eqnarray}
V_{12}(t)= \left\{
            \begin{array}{c}
              V_{{\rm max}}  \, \, \, \, \, \, \, \, \, \, \, \, \, \, \, \, \, \, \, \, \, \, \, \, \, \, \, \, \, \, \, \, \, \, \, \, \, \, \, \, \, \, \, \, \, \, \, \, \, \, \, \, \, \, \, \, \, \, \, \, \, \, \, \, \, \, \, \, \, \, \, \, \, \, \, \, \, \, \, \,\, \, \, \, \,\, \,t
              < \tau \\
              V_{23}(t-\tau) \, \, \, \, \, \, \, \, \, \, \, \, \, \, \, \, \, \, \, \, \, \, \, \, \, \, \, \, \, \, \, \, \, \, \, \, \, \, \, \, \, \, \, \, \, \, \, \, \, \,  \tau \le t<t_{{\rm p}}+\tau \\
               V_{{\rm max}}    \, \, \, \, \, \, \, \, \, \, \, \, \, \, \, \, \, \, \, \, \, \, \, \, \, \, \, \, \, \, \, \, \, \, \, \, \, \, \, \, \, \, \, \, \, \, \, \, \, \, \, \, \, \, \, \, \, \, \, \, \, \, \, \, \, \, \,\, \, \, \, \, \,   t \ge
               t_{{\rm p}}+\tau.\\
            \end{array}
          \right.
          \label{V12}
\end{eqnarray}
In order to ensure that the transport across the three-well system
is dominated by tunneling, $V_{{\rm min}}$ is greater than the
chemical potential of the BEC.

The dynamics of the BEC are determined through numerical solutions
of Eq. (\ref{GPE}) by employing the Crank-Nicholson method
\cite{Crank} to evolve the equation. The initial state at $t=0$ is
obtained via propagation in imaginary time \cite{Imaginary} subject
to the number of atoms and the potential of Eq. (\ref{potential}).
The true ground state for the system at $t=0$ consists of a
symmetric state with half the atoms in well $1$ and half in well
$3$. To initialize the system we set $\Psi({\bf r})=0$ for $z\ge 0$,
thus leaving only atoms in well $1$, as schematically shown in Fig.
1(a). Experimentally preferential loading of well $1$ can be
obtained by an initial shift in the parabolic potential
\cite{Albiez}.

For our analysis using the 1D GPE we have recast Eq. (\ref{GPE}) in
dimensionless form, in terms of harmonic oscillator units. The fundamental units of length and time are defined by $l_{{\rm
ho}}=\sqrt{\hbar/m\omega_z}$ and $t_{{\rm ho}}=1/\omega_z$
respectively, with the dimensionless interaction strength being
$g_{{\rm 1D}}=g\sqrt{m/\hbar^3 \omega_z}/(2\pi l_r^2)$,  $l_r$ is
the size of the BEC in the radial direction.

To evaluate number of atoms in wells two and three in 1D we use the
definitions,
\begin{eqnarray}
N_2(t)=\int_{-z_0}^{z_0}|\Psi(z,t)|^2 dz
\\
N_3(t)=\int^{\infty}_{z_0}|\Psi(z,t)|^2 dz.
\end{eqnarray}
In 3D, Eqs. (11) and (12) are generalized to
\begin{eqnarray}
N_2(t)=\int_{-z_0}^{z_0} \int^\infty_\infty \int^\infty_\infty
|\Psi({\bf r},t)|^2 dx dy dz \\
N_3(t)=\int^{\infty}_{z_0}
\int^\infty_\infty \int^\infty_\infty |\Psi({\bf r},t)|^2 dx dy dz,
\end{eqnarray}
which are used to define the fidelities of the process shown in Figs. 1(b) and 4(a).


\begin{thebibliography}{99}
\bibitem{note} The non-linear interaction parameter in the mode analysis, $g_m$, can be considered as being similar to the interaction parameter, $g$, in the GPE.
\end{thebibliography}
\end{document}